\documentclass[%
 reprint,
 amsmath,amssymb,
 aps,
]{revtex4-1}

\usepackage{graphicx}% Include figure files
\usepackage{dcolumn}% Align table columns on decimal point
\usepackage{bm}% bold math

\begin{document}

\title{Giant Magneto-refractive Effect in Second Harmonic Generation from Plasmonic Antennas in the Mid-infrared}

\author{Ilya Razdolski}
\email{e-mail: ilyar@science.ru.nl}
\affiliation{FELIX laboratory, Radboud University, 6525 ED Nijmegen, The Netherlands}
\author{Gaspar Armelles}
\author{Alfonso Cebollada}
\affiliation{IMM-Instituto de Microelectrónica de Madrid (CNM-CSIC), Isaac Newton 8, PTM, Tres Cantos, E-28760 Madrid, Spain}
\author{Andrei Kirilyuk}
\affiliation{FELIX laboratory, Radboud University, 6525 ED Nijmegen, The Netherlands}

\date{\today}

\begin{abstract}

Metallic nanostructures exhibit strong nonlinear-optical response at surface plasmon resonances, where the light-matter coupling efficiency is enhanced. An active modulation of this response can be realized by means of an external magnetic field. Here we utilize a nonlinear magneto-refractive effect in spintronic multilayer antennas to achieve a resonant 20\% modulation in second harmonic generation (SHG) in the mid-infrared. We discuss mechanisms of this modulation and show that it cannot be explained by an unequal enhancement of the electromagnetic field in the two spin states of the multilayer. Instead, we propose a novel contribution to the nonlinear susceptibility, which relies on the spin-dependent electron mean free path in metals. In contrast to magneto-optics in ferromagnets, our approach results in no shift of the resonance and thus ensures that the largest SHG and its strongest modulation are simultaneously observed.    

\end{abstract}

\maketitle

\section{Introduction}

Nonlinear magneto-plasmonics resides at the crossroads of multiple branches of modern photonics, aiming at utilizing elementary plasmonic excitations for enhanced, magnetic field-controlled nonlinear-optical outputs. It intrinsically relies upon surface plasmon-induced enhancement of the electric field, which facilitates nonlinear-optical processes such as, for instance, second harmonic generation (SHG). The latter is known to be sensitive to surfaces and interfaces \cite{Shen}, enabling even stronger bonds with plasmonics. Further, magnetic constituents in nonlinear plasmonic systems enable convenient non-invasive tunability by means of an external dc magnetic field, switching the magnetization and resulting in the variations of the magneto-induced SHG (or other nonlinear-optical process) output. Owing to the established relations between magnetic and non-magnetic nonlinear-optical susceptibilities, magneto-optical contrasts in SHG can reach tens of percents, or 1-2 orders of magnitude higher than in linear optics. As such, potential applications can be devised in the form of a magneto-plasmonic nonlinear-optical converter featuring high SHG efficiency and external tunability. This concept can be further extended onto an ultrafast timescale using laser-induced magnetization switching \cite{KirilyukRev10}.

Yet, conventional approaches utilizing Kerr (or Faraday) effects in various magneto-optically active media have up to date failed to address this goal. A plethora of systems have been tried, including transition metal films \cite{RazdolskiPRB13,ZhengSciRep14,ZhengOptMaterExpr15}, hybrid transition metal-noble metal multilayers \cite{TessierAPB99,PavlovAPL99,RazdolskiACS16,TemnovJO16}, noble metal-magnetic dielectric magneto-plasmonic crystals \cite{RazdolskiACS15,KrutyanskiyPRB15,ChekhovPRB16} and magnetic nanoparticles and metasurfaces \cite{KrutyanskiyPRB13,KrukACSPhot15,KolmychekOptLett16,MinhPRB18}. However, intrinsic losses introduced by the magnetic metal or relatively small nonlinear-optical susceptibility of the magnetic dielectric had a profoundly detrimental effect on the magneto-plasmonic tunability of the system. For example, in a noble metal-transition metal multilayer with a strong magnetic SHG modulation (up to 33\%) \cite{RazdolskiACS16}, the surface plasmon resonance features a decrease of the SHG output owing to the interference of multiple SHG sources at the interfaces. Recently, a fundamentally new approach to active plasmonics has been suggested \cite{ArmellesOpEx17,ArmellesACS18,ArmellesOME19}, taking inspiration from condensed matter physics, in particular, giant magneto-resistance (GMR) effect. 
%Without going too much into detail, we note that t
The key difference to the conventional magneto-optics consists in reliance of the latter on off-diagonal components of the effective dielectric tensor $\hat{\varepsilon}_{ij}$. On the contrary, in GMR systems the magnetization-induced conductivity variations are directly related to the diagonal components $\hat{\varepsilon}_{ii}$. As such, magnetic tunability of the excitation is decoupled from the losses, promising higher quality resonances in magnetic systems. 

So far this novel concept of GMR plasmonics has only been demonstrated in the linear optics. In this work, we explore the potential of GMR magneto-plasmonics in the nonlinear domain. To that end, we analyze the SHG response of plasmonic GMR antennas in the mid-infrared spectral range where the magneto-refractive effect is enhanced \cite{KravetsPRB02,MennickeJMMM06,ArmellesOME19}. We found a pronounced increase of the SHG output at the resonance, complemented by the SHG magnetic modulation (nonlinear magneto-refractive effect, NMRE) up to 20\%. The magnetic field dependencies of the SHG output clearly corroborate the results of the spectral measurements. Our work opens the new branch of nonlinear active plasmonics, extending its fundamental concepts onto a novel class of systems and bridging nonlinear (magneto-)photonics with condensed matter physics. We believe that our results will stimulate the nonlinear-optical community to look for further applications of condensed matter effects in the search for the optimized magneto-plasmonic and MRE systems.

\begin{figure}
    \centering
    \includegraphics[width=0.9\columnwidth]{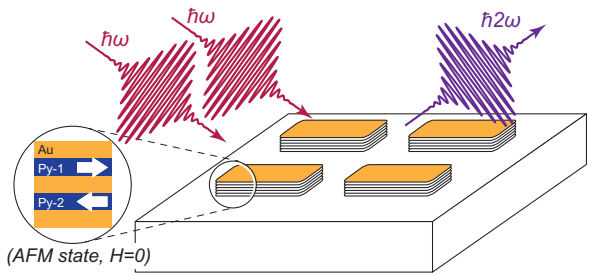}
    \caption{
    Schematic of the nonlinear-optical experiment. The inset shows the internal layered structure of the spintronic antennas. The white arrows indicate the directions of magnetization in the two Py layers in the absence of the external magnetic field (corresponding to the AFM state of the spintronic multilayer).
    }
    \label{fig:schematic}
\end{figure}

\section{Experimental}

%The antennas we studied here are intimately similar to those designed and examined elsewhere  \cite{ArmellesACS18}, demonstrating the rich potential of magneto-plasmonics in the infrared. In brief, 

We studied an array of dipole multilayer antennas \cite{ArmellesACS18} with a surface plasmon resonance at about $6.5~\mu$m wavelength. The multilayer Au/Py structure with antiferromagnetically coupled adjacent Py layers and 2 nm-thick Au spacers between them enables a magneto-refractive response of the antennas \cite{ParkinPRL94}. The spatial arrangement of the antennas played no role for the optical properties in the interesting spectral region due to the small array period of about $2~\mu$m. Figure~\ref{fig:schematic} illustrates the schematics of the nonlinear-optical experiment as well as the antennas inner structure. We employed the FELIX free-electron laser as a source of fundamental radiation (about $5$~mJ energy in the macropulse, 10 Hz repetition rate, and 25~MHz repetition rate of the picosecond micropulses within every macropulse) tunable in the wide range of wavelengths in the infrared ($5-100~\mu$m). Further details about FELIX can be found elsewhere \cite{KnippelsPRL99,KirilyukPRB00a}. By means of a $2''$ parabolic mirror the beam was focused onto the sample into a spot of about $300~\mu$m in diameter.  Intrinsically generated higher harmonics in the fundamental beam were cut off by the LWP $4.5~\mu$m (Spectrogon) filter whereas in the reflected SHG beam, the fundamental component was filtered out using the top-hat $4~\mu$m filter (Edmund) and the 3~mm-thick CaF$_2$ plate (Andover). The polarization of the fundamental radiation incident at about 30~degrees was controlled by two wire-grid polarizers (Edmund) mounted on rotational stages. The filtered SHG radiation was registered by a nitrogen-cooled MCT photodetector. Magnetic field up to 150~Oe in the in-plane direction was applied by means of an electromagnet.

\begin{figure}[t]
    \centering
    \includegraphics[width=0.95\columnwidth]{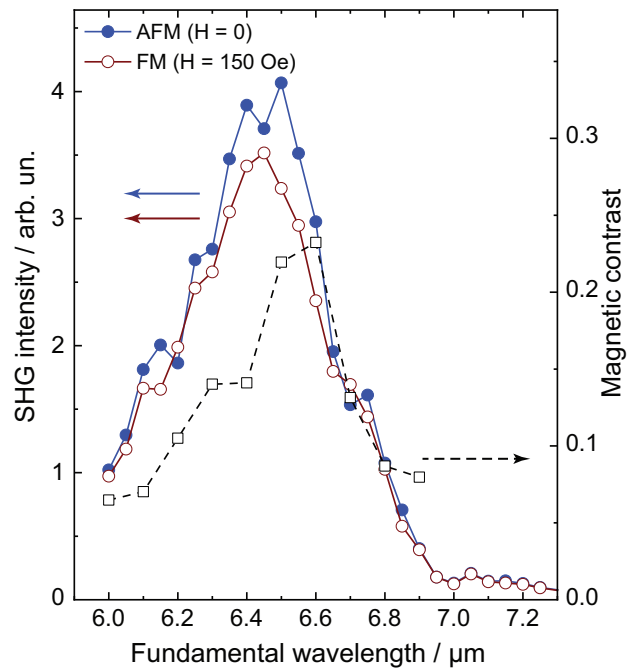}
    \caption{
    Experimental SHG spectra measured on spintronic antennas in the vicinity of the surface plasmon resonance (left axis). The full blue and empty red circles correspond to the AFM and FM states of the multilayer, respectively. The empty gray squares indicate the spectral dependence of the SHG magnetic contrast $\rho$ given by Eq.~\ref{contrast} (right axis). 
    }
    \label{fig:spectra}
\end{figure}

In previous works \cite{ArmellesACS18, ArmellesNanophot19} it was shown that these antennas exhibit a surface plasmon resonance featuring a pronounced enhancement of the electric field. As such, it is natural to expect a peak in the SHG output spectrum at around the same wavelengths. Indeed, Fig.~\ref{fig:spectra} illustrates the experimental SHG spectra measured with the p-polarized incident fundamental radiation. A clear broad maximum is seen, centered at about $6.5~\mu$m, while similar measurements with the s-polarized fundamental radiation showed no SHG output beyond the noise level (not shown). It is also seen that the SHG spectra in zero magnetic field (with antiparallel orientation of magnetization in the two Py layers) and in saturating magnetic field (where the magnetizations are aligned parallel) are different, indicating significant magneto-induced SHG contribution. To quantify this difference, we introduce the magneto-optical SHG contrast $\rho$:

\begin{equation}
\label{contrast}
    \rho=\frac{I^{2\omega}_0-I^{2\omega}_H}{I^{2\omega}_0}
\end{equation}

\begin{figure}[t]
    \centering
    \includegraphics[width=0.95\columnwidth]{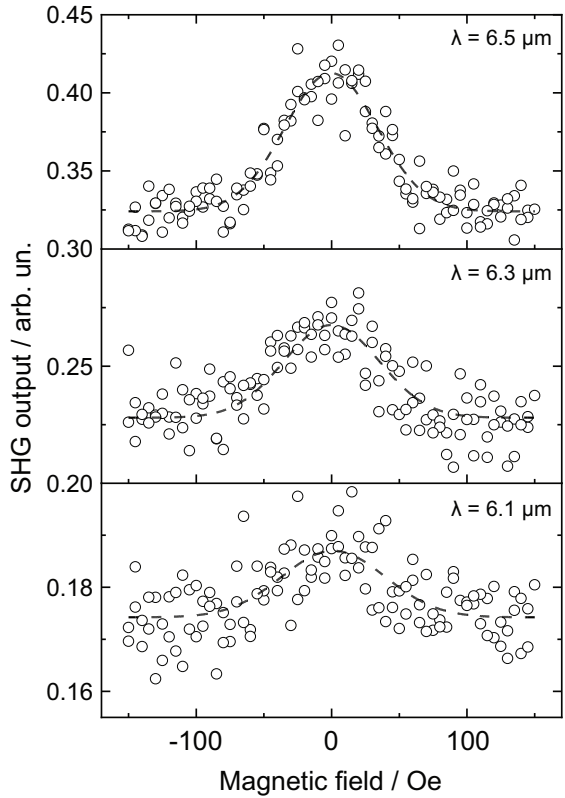}
    \caption{
    SHG output dependence on the external in-plane magnetic field exemplified for three fundamental wavelengths in the vicinity of the surface plasmon resonance of the spintronic antennas. The dashed lines are guides to the eye.
    }
    \label{fig:magfield}
\end{figure}

The open squares in Fig.~\ref{fig:spectra} illustrate the variations of $\rho$ across the resonant SHG peak. We shall discuss the apparent increase of the contrast towards longer wavelength a bit later, and for now just highlight the observation that $\rho$ reaches $15-20\%$ while the SHG output is still reasonably large. 
%Because the signal-to-noise ratio in the data exemplified above is not impressively large, precise determination of the SHG magnetic contrast values is difficult. 
To take a closer look at the magneto-induced SHG contribution, we performed systematic measurements of the SHG output as a function of the in-plane magnetic field at various wavelengths. Typical datasets are illustrated in Fig.~\ref{fig:magfield} for a few fundamental wavelengths where the nonlinear-optical magneto-refractive effect is seen at its best. The symmetrical shape of the curves indicates an even in magnetization effect, consistent with the magneto-refractive mechanism.

\section{Discussion}

These results confirm large (up to $20\%$) magneto-refractive modulation of the SHG output at the surface plasmon resonance where the SHG efficiency is enhanced. This observation is in a striking contrast with conventional magneto-plasmonic systems \cite{MaccaferriPerspective} where large magnetic contrasts are accompanied by minima in the total SHG output. Intuitively, this picture can be understood as a result of the resonance shift which is very characteristic for the conventional magneto-optical effects originating in non-diagonal components of the dielectric tensor.
%manifestation of the Kramers-Kronig relations. Indeed, spectra of the diagonal and non-diagonal (responsible for magneto-optics) dielectric tensor components are shifted with respect to each other. 
%Because conventional magneto-optics relies on onto non-diagonal components of the dielectric tensor, this leads to the 
On the contrary, the aforementioned reliance of the magneto-refractive approach onto diagonal components of the dielectric tensor ensures that magneto-refractive plasmonic systems have a great potential for ultimately addressing the principal issue of nonlinear magnetoplasmonics.
%
%Due to aforementioned reliance of the magneto-refractive approach onto diagonal components of the dielectric tensor, magneto-optics and Joule losses are decoupled, unlike in the conventional approaches which suffer from their inability to accompany enhanced SHG yield with large magneto-optical effect. 
%Ultimately addressing the principal issue of nonlinear magnetoplasmonics,
The experimentally observed coexistense of a strong SHG output and its large magnetic modulation illustrates high promise of the entire magneto-refractive concept for future nonlinear plasmonics.

%We note the apparent lack of magnetic hysteresis in the datasets shown in Fig.~3, contrary to the . 

The main driving force behind the observed increase of the SHG output at the SP resonance is the enhancement of the local electric field. Yet, the most important result is pertinent to the magneto-refractive modulation of the already enhanced SHG output. To understand this large modulation, we first tried to model the obtained results. The SHG output in the AFM and FM states is given by the product of the corresponding nonlinear susceptibility and the square of the local field enhancement,

\begin{equation}
\label{intensity}
    I^{SHG}_i\propto|\chi^{(2)}_iL_i^2|^2, i={\rm FM}, {\rm AFM}.   
\end{equation}

%, in particular, its in-plane projection $E_{\parallel}$. 

\begin{figure}[t]
    \centering
    \includegraphics[width=0.9\columnwidth]{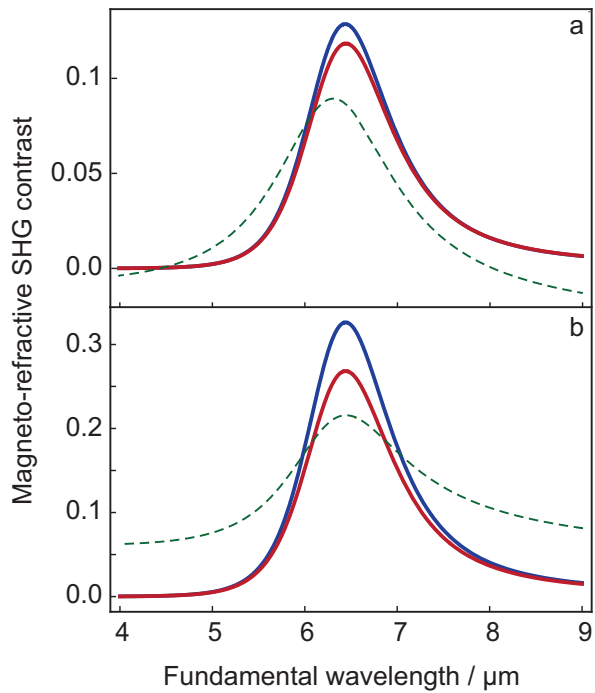}
    \caption{
    Simulated spectra of the SHG output (red and blue) in the FM and AFM states of the multilayer, respectively, and SHG magneto-refractive contrast, obtained without (a) and with (b) the phenomenological magneto-refractive contribution to the nonlinear-optical susceptibility.
    }
    \label{fig:simulations}
\end{figure}

We used the effective dielectric functions from Ref.~\cite{ArmellesOME19} to calculate the local field factors $L_i$ in the two states of the spintronic spin valve. Interestingly, the enhancement degree is unequal in the FM and AFM states: the model predicts about 1.5\% stronger enhancement $L(\lambda)$ of the electric field in the AFM state. Thus the experimentally obtained higher SHG output in the AFM state is in a qualitative agreement with the magneto-refractive mechanism of the SHG modulation. Further, the magneto-refractive variations of the dielectric function $\varepsilon$ upon the AFM-to-FM transition result in the similar variations of the nonlinear-optical susceptibility $\chi^{(2)}$. At metallic surfaces, the conventionally used models in the spirit of the Rudnick and Stern approach \cite{RudnickStern} assume $\chi^{(2)}\propto\varepsilon-1$. The results of this purely magneto-refractive model including both factors ($\chi^{(2)}$ and $2$) are summarized in Fig.~\ref{fig:simulations},a. The SHG peak at about $\lambda=6.5~\mu$m is qualitatively well reproduced, yet the magnetic contrast does not exceed $8\%$. As such, the resonant field enhancement together with the magneto-refractive variation of dielectric function does not explain large (up to $20\%$) SHG magnetic contrast values. 

In order to get a deeper insight into the mechanism of the nonlinear-optical modulation of the SHG output at the resonance, we recall the inherently surface origin of the dipole second order nonlinearity in metals. Indeed, owing to the inversion symmetry in the bulk, in the dipole approximation non-zero $\chi^{(2)}$ originates in the electron scattering at surfaces and interfaces. It can thus be conjectured that effective $\chi^{(2)}$ should depend on the electron scattering rate $\gamma$ (as well as the related electron mean free path $l_e$). Similar conclusions have been reached in previous works \cite{LiebschPRB89,SchaichPRB88}. In the FM (open) state of the spintronic valve, electrons can freely traverse the multilayer, thus having a larger mean free path than in the AFM (closed) state. In other words, transition to the AFM state results in additional effective interfaces acting as new SHG sources. Considering two AFM-coupled sublattices of magnetic Py layers with magnetizations $\vec{M}_1$ and $\vec{M}_2$, we can thus expand the nonlinear susceptibility in a series:

\begin{equation}
\label{expansion}
    \chi^{(2)}(l_e) = \chi^{(2)}(\vec{M}_1\cdot\vec{M}_2)\approx \chi^{(2,nm)}+\chi^{(2,m)}\vec{M}_1\cdot\vec{M}_2.
\end{equation}

Here only the first-order term remains, and the scalar product of the magnetizations of the two sublattices indicates the dependence on the FM or AFM state. Note that both magnetizations only enter in the scalar product, and the tensor symmetry of the second term (magnetic, $\chi^{(2,m)}$) is exactly the same as the first one (non-magnetic, $\chi^{(2,nm)}$). Once higher-order terms are omitted, and only two states of the spin valve are considered, the expression for the nonlinear susceptibility can be further simplified to $\chi^{(2)}(1\pm r)$, where $\chi^{(2)}\propto\varepsilon-1$ is the common prefactor, $r$ is the (complex) ratio of the magnetic and non-magnetic contributions, and the sign is determined by the FM or AFM state of the spin valve. Further Eq.~\ref{intensity} can be employed to calculate the SHG spectra in the two states as well as the magneto-refractive contrast $\rho$ using the amplitude and the phase of $r=|r|e^{i\varphi}$ as free parameters. 

This approach allows to get a much better agreement with the experimental data, as shown in Fig.~\ref{fig:simulations},b. Magnetic contrast peaks at the resonance, reaching about $20\%$ at $|r|=0.05$, i.e. the second term in Eq.~\ref{expansion} is only $5\%$ of the first one. Together with the phase $\varphi\approx0.9\pi$, this ensures the off-resonant non-zero contrast visible in the calculations but not in the experiment due to the low SHG output. Owing to the resonant response of the antennas, the SHG output peaks simultaneously with the magneto-refractive contrast.
%, taking advantage of the stronger field enhancement in the AFM state of the spin valve. 
At the resonance, the in-plane projection of the electric field $E_{\parallel}$ is enhanced, facilitating the nonlinear contribution associated with the $\chi^{(2)}_{\perp\parallel\parallel}$ tensor component. The scattering rate of electrons moving along the $\perp$ direction at the double frequency is thus controlled by the external magnetic field capable of switching between the FM and AFM states, highlighting the nonlinear-optical features of the magneto-refractive effect in plasmonic antennas. Yet, thorough modeling of the magneto-refractive SHG response of the resonant plasmonic antennas requires accounting for all three non-zero $\chi^{(2)}$ components ($\perp\parallel\parallel$, $\perp\perp\perp$ and $\parallel\perp\parallel$) with their amplitudes and phases, allowing too many free parameters for a meaningful fit routine. 

It might seem confusing that the first non-neglected magnetic term in Eq.~\ref{expansion} is quadratic in total magnetization $M$ in the FM state (since $M_1=M_2=M/2$). Conventional approaches in nonlinear magneto-optics consider odd in magnetization terms \cite{RuPinPan}. The physical mechanism of these terms, corresponding to the non-diagonal terms of the dielectric tensor $\varepsilon$ in linear magneto-optics, is pertinent to the spin-orbit coupling.  Yet, it is generally accepted that in the infrared spectral range the magnitude of these terms becomes negligibly small, together with the ratio $\omega/\omega_p$. Here $\omega_p$ is the plasma frequency, usually on the order of a few eV in plasmonic metals. In contrast, the "magnetic" $\chi^{(2,m)}$ term considered here has a completely different origin. In fact, its magnetic nature is only ensured by the spintronic control of the effective electron scattering rate in the multilayer antennas. This entirely different physical mechanism explains the even parity of this term with respect to the total magnetization as well as its sizeable magnitude in the infrared range where $\omega/\omega_p$ is very small, and the optical response of metals is dominated by the intraband transitions of conduction electrons.

\section{Conclusions}

To summarize, we have demonstrated for the first time the magneto-refractive effect in SHG from plasmonic antennas consisting of spintronic Au/Py multilayers. Application of a weak magnetic field results in a transition from the FM to the AFM state, accompanied by about 20\% modulation of the SHG output. The effect has a clear resonant nature and is attributed to the excitation of the localized surface plasmon mode in the mid-infrared spectral range. The experimental results are modeled within a SHG magneto-refractive approach taking into account variations of the second-order effective susceptibility upon switching of the spin valve. Building a bridge between plasmonics and spintronics, our results demonstrate the high promise of a whole new class of systems for future magneto-photonic applications. 

\acknowledgments

We acknowledge financial support from MINECO through projects AMES (MAT 2014-58860-P), Quantum Spin Plasmonics (FIS2015-72035-EXP) MIRRAS (MAT2017-84009-R), and Comunidad de Madrid through project SINOXPHOS-CM (S2018/BAA-4403). We acknowledge the service from the MiNa Laboratory at IMN and funding from MINECO under project CSIC13-4E-1794 and from CM under project S2013/ICE- 2822 (Space-Tec), both with support from EU (FEDER, FSE).

%\suppinfo

\bibliography{source}

\end{document}